# Photocurrent imaging and efficient photon detection in a graphene transistor

Fengnian Xia[1]*, Thomas Mueller[1], Roksana Golizadeh-mojarad[2], Marcus Freitag[1], Yu-ming Lin[1], James Tsang[1], Vasili Perebeinos[1], and Phaedon Avouris[1]*

[1]IBM Thomas J. Watson Research Centre, Yorktown Heights, New York 10598

[2]School of Electrical and Computer Engineering, Purdue University, West Lafayette, Indiana, 47906

*email: fxia@us.ibm.com, avouris@us.ibm.com

**Abstract.** We measure the channel potential of a graphene transistor using a scanning photocurrent imaging technique. We show that at a certain gate bias, the impact of the metal on the channel potential profile extends into the channel for more than 1/3 of the total channel length from both source and drain sides, hence most of the channel is affected by the metal. The potential barrier between the metal controlled graphene and bulk graphene channel is also measured at various gate biases. As the gate bias exceeds the Dirac point voltage, $V_{Dirac}$, the original p-type graphene channel turns into a p-n-p channel. When light is focused on the p-n junctions, an impressive external responsivity of 0.001 A/W is achieved, given that only a single layer of atoms are involved in photon detection.

Graphene has recently attracted strong attention due to its unique properties and possible applications in nanoelectronics[1-3] and nanophotonics[4]. However, the role of source and drain metal contacts and their impact on the channel potential profile deserve more attention. Here we experimentally studied the channel potential profile of graphene transistors at various gate biases using a scanning photocurrent imaging technique[5-9]. This technique[5-9] has previously been used to study the internal electric field in nanoscale devices. In this approach, the nanostructure is locally excited by photons. A photo-generated current image is then produced as the laser spot is scanned across the device. Since the lifetimes of the photon-generated carriers are very short[10, 11] (in the ps range), the current resulting from carrier diffusion is negligible. Hence, the generation of photocurrent is attributed to the presence of a local, in-plane electric field within the device and the magnitude of the photocurrent is a direct measure of the strength of this field[5]. Here, we use this approach to study the internal electric field of functioning graphene transistors at various gate biases. We also supplement our experimental studies with quantum transport simulation using the nonequilibrium Green's function (NEGF)[12-16] approach self-consistently coupled to a Poisson solver[15] for treating the electrostatics. The results obtained using this optical approach shed light on important issues in graphene transistor such as the minimum conductance[17] and the asymmetric conduction behavior for electrons and holes[18].

The inset of Fig. 1a shows a typical graphene transistor of the type studied here. The graphene channel in this device is 0.62 μm wide and 1.45 μm long. The fabrication process is presented in the "supporting information". Device characterizations were performed at ambient conditions. A typical electrical transport characteristic at a drain bias of 1 mV is shown in Fig. 1a. The minimum current is achieved at a gate bias $V_{Dirac}$ of +30 V, due to trapped negative charges in the $SiO_2$ substrate. These charges field dope the graphene p-type. Raman scattering results obtained from the graphene transistor channel shown above are plotted in the inset of Fig. 1c. The scattering intensity of the 2D mode is about two times as large as that of the G mode, confirming that the graphene under investigation is indeed a single-layer[19].

The graphene transistor was first located using the optical reflection image (Fig. 1b) before the scanning photocurrent imaging measurement (described in the supporting information). Here, only the source and drain metal contacts are visible since they reflect light effectively. A photocurrent image measured at $V_D=V_S=V_G=0$ V is plotted in Fig. 1c. The photocurrent image exhibits two major features: First, two peaks of about 15 nA appear when the excitation laser spot is located near the metal contacts, indicating the presence of strong local electric fields close to the contacts. Second, the photocurrent signs (indicating the current flow direction) on each side of the contact are opposite and the photocurrent image possesses almost perfect inversion symmetry, which implies a mirror symmetry of the electric field (or of the potential profile) within the channel since the current flowing towards (or away from) the identical source and drain contacts will result in photocurrents in an external circuit with opposite flow directions.

In order to study the gate dependence of the channel potential profile, we performed photocurrent line scans along the center of channel as shown by the dashed line in Fig. 1c at gate biases from -50 V to +60 V. Fig. 2a shows a schematic of the laser spot and the graphene transistor in this scan. Seven photocurrent response curves are plotted in Fig. 2b. The peak position and peak magnitude of the photocurrents are extracted and plotted in Fig. 2c and d, respectively. When gate bias is between -50 to 0 V, the two major features of the photocurrent at $V_G=0$ V remain unchanged. However, the peak magnitude and peak position depend moderately on the gate bias as shown in Fig. 2c and d.

When the gate bias $V_G$ is between +10 and +15 V, the photocurrent peak magnitudes are reduced by about 75% when compared with those at $V_G=0$ V as shown in Fig. 2b and d, indicating a decrease in the magnitude of the local in-plain electric field. At the same time, the photocurrent peak position moves towards the center of the channel by around 450 nm in both source and drain sides as shown in Fig. 2b and 2c. Most importantly, at $V_G=+15$ V which is 15 V smaller than the Dirac point voltage, $V_{Dirac}$, the polarity of the photocurrent starts to flip, implying the reversal of the band bending direction. The photocurrent peak magnitude increases moderately between +20 and +60 V and the peak position of the photocurrent moves back towards the contact, as shown in Fig. 2c and d. However, in contrast to the case at $V_G=-50$ V, where the maximum photocurrent is generated right at the contacts, at +60 V, the photocurrent maximum is still about 200 nm away from the contact and about 40% larger than that obtained at $V_G=-50$ V.

The evolution of the photocurrent under various gate biases can be explained by examining the band profiles within the channel. Fig. 3a shows a schematic drawing of half of the graphene transistor. The following panels (Fig. 3b to 3e) show the approximate band profiles at various gate voltages inferred from photocurrent line scans (Fig. 2b) and the transfer characteristics (Fig. 1a) measurements. The exact shape of band profile cannot be determined accurately due to rather large excitation laser spot (550 nm). The dashed blue curve denotes the Fermi level and the solid black line denotes the energy at the graphene Dirac point, respectively. The graphene transistor can be divided into three segments. On the left is the metal dominated region (segment I) where the graphene is covered and controlled solely by the metal. We assume here that the interactions with the metal and the dielectric do not significantly perturb the basic electronic structure of the graphene, but simply broaden the bands[20, 21]. However, the metal does dope the graphene under it[20], while the gate field has a negligible impact on this segment. The bulk graphene channel (segment III) is on the right and the carrier density within this segment is controlled by the gate. A simple capacitor model[4] can be applied to determine the carrier density (characterized by $\Delta E$ as shown in Figs. 3b to 3e). A transition region (segment II) is located in between the other two and its carrier density is affected by both the metal and gate bias. In the steady state, the Fermi level is fixed and gate-independent since it is controlled by the source and drain contacts ($V_S = V_D = 0$ V). Varying the gate bias changes the doping of the bulk graphene channel and also causes a charge redistribution, leading to a variation of the potential profile and hence of the photocurrent.

The doping of the graphene under metal contacts, which is characterized by $\Delta\phi$ shown in Figs. 3b to 3e, is independent of gate bias. The band-bending direction in the transition region (segment II) is determined by $\Delta\phi$ and $\Delta E$. The polarity of the photocurrent flips at a positive gate voltage, indicating that reduction of ΔE results in the reversal of the band bending direction. Hence originally at $V_G$=0 V, we have ΔE > Δϕ as shown in Fig. 3b.

The minimum conductance occurs at $V_G$=$V_{Dirac}$=+30 V (Fig. 1a), and a ΔE of around 140 meV is estimated[4] at $V_G$=0 V. The maximum photocurrent is measured when the laser spot moves away from the contact by about 100nm, from which we can infer that the band bending extends into graphene by at least 100 nm. As $V_G$ decreases from zero to negative voltages, the p doping of the bulk graphene channel is enhanced, leading to an increased ΔE, while the transition region (segment II) width decreases. This is analogous to a conventional metal-semiconductor or semiconductor-semiconductor interface in which the depletion region decreases as the doping level increases[22]. The impact of a decreased gate bias on the photocurrent is twofold. The local electric field increases since the band bending is enhanced, which is expected to enhance the photocurrent. On the other hand, the width of the band-bending region (segment II in Fig. 3b and c) is reduced, leading to a smaller effective light absorption area and a decline in photocurrent. The experimental observations show a rather moderate increase and a clear trend of saturation in photocurrent when $V_G$ decreases from 0 to -50 V, a result of these two opposing influences.

At $V_G$=+15 V, ΔE is approximately equal to Δϕ, resulting in a rather flat potential profile as shown in Fig. 3d. The calculated height, Δϕ, using the approach mentioned above is about 95 meV. Since Δϕ is gate independent, we can infer that the energy barrier between the contact metal controlled graphene and bulk graphene is around 45 meV (ΔE−Δϕ) at $V_G$=0 V. Increasing the gate bias $V_G$ to +45 V leads to the potential profile shown in Fig. 3e. The peak photocurrent is enhanced from 5 nA at $V_G$=+15 V to 25 nA at $V_G$=+45 V. One interesting observation is that when $V_G$ is greater than $V_{Dirac}$, the p type graphene channel turns to p-n-p type channel and a maximal photocurrent response is observed when light is focused on the p-n junction.

Another interesting observation which can be explained by the internal electric field distribution is the photocurrent resulting from internal photoemission/thermal injection[5] from metal to graphene when the laser spot is focused on the metal contacts. As shown in Fig. 2a and 2b, when laser spot is located at a position of 2 µm (i.e. inside the electrode), at which the graphene channel is hardly illuminated by the light, photocurrents of around +7 and -4 nA are observed at a gate bias of -40 and +60 V, respectively, but are almost negligible at +15 and +20 V. To generate a current, carriers produced by internal photoemission/thermal injection[5] also require the presence of an internal in-plane electric field within the graphene channel; hence the photocurrent resulting from the internal photoemission/thermal injection is negligible at $V_G$ of +15 V. Moreover, the magnitude of this photocurrent also depends on the separation of the internal electric field and carrier generation locations. When gate bias is between +20 and +60 V, the internal electric field within the graphene channel is located at least 200 nm away from the metal contacts and hence the photocurrent due to internal photoemission is much smaller than that observed at negative gate biases.

To support the above analysis of the experimental data, we have also performed a full real-space quantum transport simulation using the nonequilibrium Green's function (NEGF)[12-16] approach self-consistently coupled to a Poisson solver for treating the electrostatics. This allows us to calculate the charge distribution and electrostatic potential in the device. Metal induced gap states[23-24] are taken into account. This simulation allows us to calculate the charge distribution and electrostatic potential in the device, describing the change in the graphene band profile due to the spatially varying electrostatic potential.

The electronic properties of the graphene channel are described using a tight-binding framework with one $p_z$ orbital per carbon atom and a coupling t=2.7 eV between nearest neighbor atoms. From NEGF we obtain spatial surface charge distribution across the channel using:

$$\rho_s = \frac{e}{2\pi} \int dE G^n, \qquad (1)$$

where e is the electron charge and $G^n = G\Sigma^{in}G^+$.

The Green's function G is calculated by solving:

$$G = [(E - eU)I - H_0 - U_B - \Sigma_S - \Sigma_D]^{-1}, \qquad (2)$$

where $H_0$ represents the channel Hamiltonian using a $\pi$-orbital nearest neighbor tight binding model, $U_B$ (140 meV) is the channel doping potential barrier extracted from Dirac point gate voltage, U is the electrostatic potential on the graphene channel, and $\Sigma_S$ ($\Sigma_D$) includes the interaction of semi-infinite source (drain) contact with the channel, respectively. In general,

$$\Sigma_{S,D} = \tau_{S,D} g_{S,D} \tau_{S,D}^+, \qquad (3)$$

and

$$\Sigma^{in} = -2(f_S \Sigma_S + f_D \Sigma_D),  \quad (4)$$

where $\tau_S$ ($\tau_D$) represents the coupling between the channel and source (drain) contact, $g_S$ ($g_D$) is the surface Green's function for the semi-infinite source (drain), and $f_S$ ($f_D$) is the source (drain) Fermi function. In the experiment the source and drain are grounded, hence $f_S$ ($f_D$) = 0.5 at E=$E_F$. The self-energy $\Sigma_S$ ($\Sigma_D$) accounts the physical properties of the contacts. The graphene electrodes are controlled mainly by the metal and the gate bias has negligible impact on them. Due to the band lineup at the interface of the metal and graphene electrodes (as shown in Fig 3b to 3f), a fixed potential-energy barrier $\Delta\phi$ (~95meV) is introduced in the graphene electrode. Note that this value is extracted from the flat band photocurrent measurement at a gate bias $V_G$=+15 V (Fig. 2b). The surface Green's function $g_S$ ($g_D$) for the graphene electrode is obtained from the Hamiltonian for the isolated electrode ($H_{electrode}$) using

$$g_S = (E + i\eta - H_{electrode} - \Delta\phi)^{-1},  \quad (5)$$

which is evaluated by the recursive Sancho-Rubio method[16]. We consider the semi-infinite graphene electrodes as the contacts with a broadened density of states (characterized by $\eta$=0.05 eV) due to the contact with metal. The spatially varying electrostatic potential across the channel (U) is calculated by solving Poisson's equation for the device. Because the width of the channel is rather wide (620nm), we assume a uniform charge and potential distribution in the z-direction (Fig. 3a). This assumption reduces the three-dimensional (3D) Poisson's equation to a two-dimensional partial differential equation in x- and y-directions. We then use a finite-difference method to solve this two-dimensional equation numerically. A grid spacing of Δ=1.7 nm in both x- and y- direction is used.

The simulation procedure starts from an initial guess for the potential U and subsequently finds the solution for the system Green's function G (Eq. 2). Then the charge distribution $\rho_S$ on the graphene channel (Eq. 1) is calculated. Afterwards the potential profile across the channel (U) is re-calculated based on Poisson's equation. This process is iterated until both the potential and charge distributions converge. Finally the in-plane electric field ($E_x$) is obtained from the quasi Fermi level inside the channel and plotted in Fig. 4. Our simulations (Fig. 4) capture the photocurrent signatures in Fig. 2b remarkably well. The discrepancies between the simulation and experimental results are mainly due to the rather large laser spot (~550nm) used to measure the photocurrent, however, our simulation shows the same electric field directions and trends as observed experimentally. The calculated electric fields for various gate biases support the validity of the band diagrams shown in Fig. 3b to 3e and the explanations provided above.

A most interesting aspect of the theoretical investigation involves the role of the metal induced gap states[23, 24] (MIGS). If we ignore the MIGS effect, we do not obtain electric field inside the channel under flat band conditions ($V_G$=+15 V). However, by including the MIGS effect an electric field can be generated. The magnitude of this electric field depends on η in the electrodes (see Eq. 6) which accounts for the amount of MIGS inside the channel. The MIGS penetrate into the channel and increase the channel density of states (DOS). The spatial variation of the DOS leads to the spatial variation of the quasi

Fermi level in the graphene channel, leading to a local in-plain electric field. When the Dirac point energies of the graphene under metal contact and within the bulk channel are aligned, the broadened and non-zero DOS of metal controlled graphene at the Dirac point energy can have significant impact on the behavior of the graphene channel, which in the free state has zero DOS. At other biasing conditions, MIGS do not have substantial effect on the electric fields inside the channel since the DOS inside the channel is now much higher compared to that at the flat band condition.

Regarding the magnitude of the photocurrent, a maximum photocurrent of ~30 nA is achieved when the laser spot is focused on the p-n junction at the drain side at a $V_G$ of +60 V. We notice that at this gate bias, the source and drain photocurrents show a slight asymmetry. Given the incident power of ~ 30 µW, the external photon responsivity of this graphene device is 0.001 A/W at a wavelength of 632.8nm (corresponding to a 0.2% efficiency), an impressive value given that only a single layer of atoms (~0.3 nm thick) are involved in photo-detection. If silicon was used instead as a photodiode to detect light with this wavelength, about 3 nm of silicon will be needed to achieve similar responsivity even assuming that 100% of the incident light is coupled to the silicon photodiode and 100% of the absorbed light is converted into current. Moreover, if we assume that every incident photon generates an electron-hole pair in graphene, about 10% of the photon-generated electron-hole pairs are converted to current given that a single-layered graphene can absorb around 2% of the incident light[25]. This limited conversion efficiency is mainly due to the limited length of the photo-detection graphene channel in which strong local electric field exists. The conversion efficiency can be improved by introducing split gates in graphene transistors to create p-n junction so that the length and electric field magnitude of the photo-detection graphene channel can be adjusted separately. Together with the high mobility of the carriers, graphene is hence a promising material for high frequency nanophotonic applications if successful integration of graphene with high confinement optical waveguide structures and a proper wiring scheme for high frequency operation can be realized[4, 26-29].

In summary, we determined the potential profile within a p-type graphene transistor using a photocurrent imaging approach. Potential steps between the metal-controlled graphene and the bulk graphene are estimated to be around 45 meV for Ti-Pd-Au contacts at zero gate bias. When the Dirac point energy of the bulk graphene channel is aligned with that of the metal-controlled graphene, the impact of the metal on the graphene channel can extend beyond the metal for more than 450 nm. In the n-type conduction regime, the graphene close to the contacts stays p-type, leading to p-n-p channel. This might explain the experimentally extracted asymmetric conduction behavior in the p- and n- branches of graphene transistors[18]. Since the graphene sheet does not uniformly switch polarity, the minimum conductivity can also be affected, depending on metal work functions and channel geometries. Most importantly, we show that graphene might have important applications in nanophotonics due to its strong interaction with light at a wide range of wavelengths[25] and its high carrier mobility.

**Acknowledgement**

The authors thank Z. Chen for helpful discussions and B. Ek for technical support. One of the authors (T.M.) acknowledges financial support by the Austrian Science Fund (FWF; Schrödinger fellowship). Correspondence and request for materials should be addressed to F.X. (fxia@us.ibm.com) and P.A. (avouris@us.ibm.com).

**Supporting Information Available:** Detailed device fabrication and photocurrent measurement processes. This material is available free of charge via the Internet at http://pubs.acs.org.

**Figure captions**

**Figure 1 Photocurrent imaging of a graphene transistor**

**a,** Electrical transport characteristic of a graphene transistor (drain current vs gate bias) at a drain bias of 1 mV. Minimum conductance is achieved at a gate bias of +30 V. Inset: scanning electron micrograph of the graphene transistor. **b**, Optical reflection image of the graphene transistor shown in the inset of Fig. 1a. **c**, A scanning photocurrent image of the graphene transistor shown in Fig. 1a at the biasing condition of $V_S=V_D=V_G=0$ V. The incident excitation laser power is about 30 µW and the wavelength is 632.8 nm. Scale bars in **b** and **c**, 2 µm.

**Figure 2 Gate variable photocurrent profiles**

**a**, Schematic of the photocurrent line scan showing the source and drain metal contacts, the graphene channel, and the excitation laser spot. **b**, Photocurrent line scan profiles at gate biases of -40, -20, 0, +15, +20, +30 and +60 V, respectively (from panel 1 to 7). **c** and **d**, Peak photocurrent position (**c**) and magnitude (**d**) around source and drain contacts as a function of the gate bias. The data were extracted from photocurrent line scan results shown in Fig. 2b. A maximal photon responsivity of 0.001 A/W is realized at a gate bias $V_G$ of +60 V.

**Figure 3 Surface potential profiles of the graphene transistor**

**a,** Schematic of the cross section of the graphene transistor. Only half of the device is shown since the device is symmetric. **b**, **c**, **d**, and **e**, Surface potential profiles inferred from the photocurrent line scans at gate biases of 0, -20, +15, and +45 V, respectively.

Segments I, II, and III are the metal-controlled graphene, the transition region affected by both the metal and the back gate, and the bulk graphene region controlled by the back gate solely. The blue dashed line denotes the Fermi level, the black solid line represents the energy at the Dirac point of the graphene, the elongated pink rectangle stands for the graphene, and the yellow rectangle is the metal contact. The red crosses show the linear dispersion around Dirac point. At $V_G > V_{Dirac}$, the graphene channel close to the contact remains p type although the bulk graphene channel turns into n type as shown in **e**. Here, $\Delta\phi$ is analogous to the Schottky barrier height in conventional metal-semiconductor contact.

**Figure 4 Calculated local in-plane electric field profiles at various gate biases**

Calculated electric field profiles within the graphene channel at gate biases of -40, -20, 0 +15, +20, +30, and +60 V, respectively. The simulation is based on NEGF approach self-consistently coupled to a Poisson's equation. Metal-induced gap states are taken into account. The results capture the signature of the photocurrents shown in Fig. 2b very well.

Figure 1

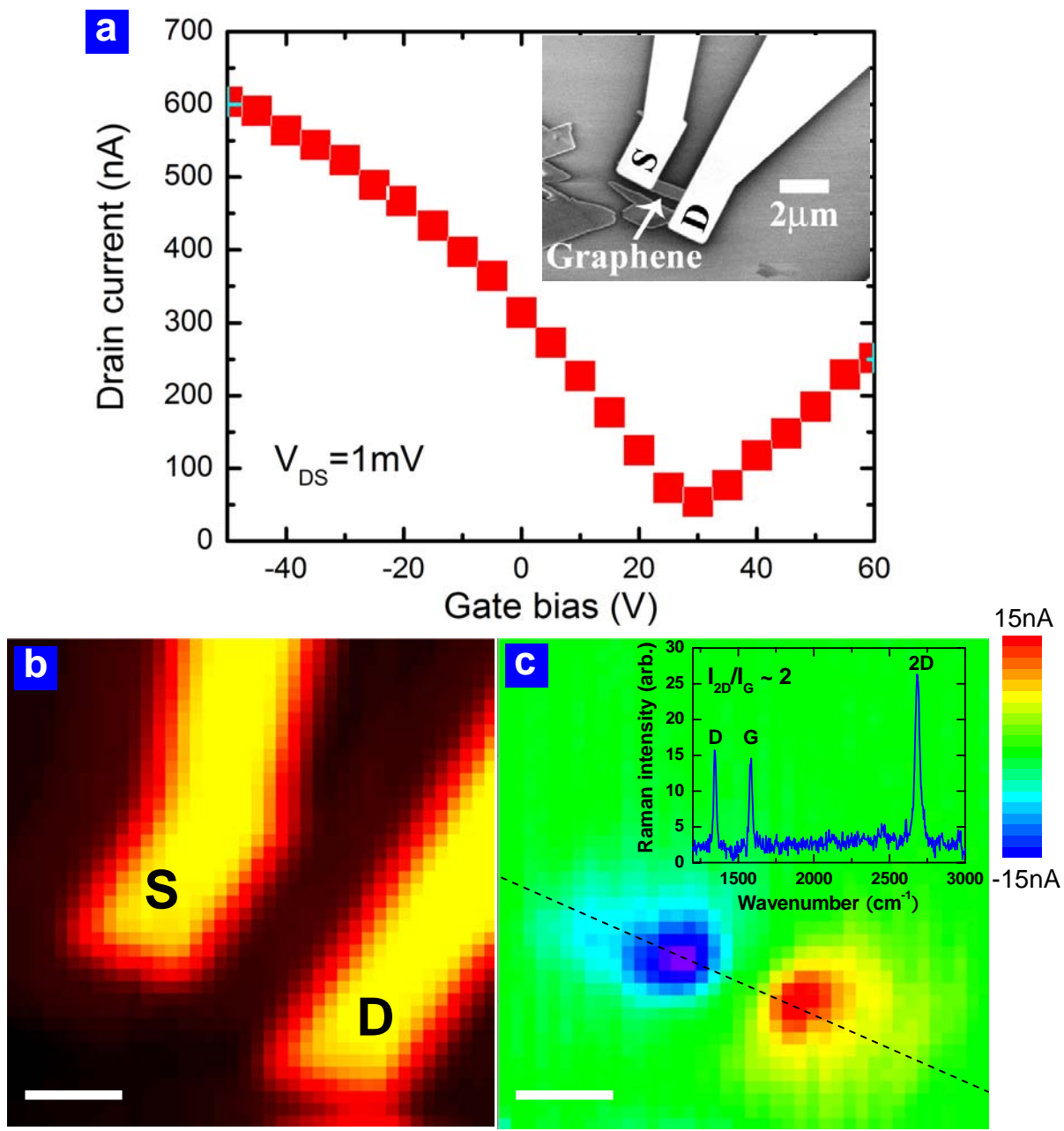

Figure 2

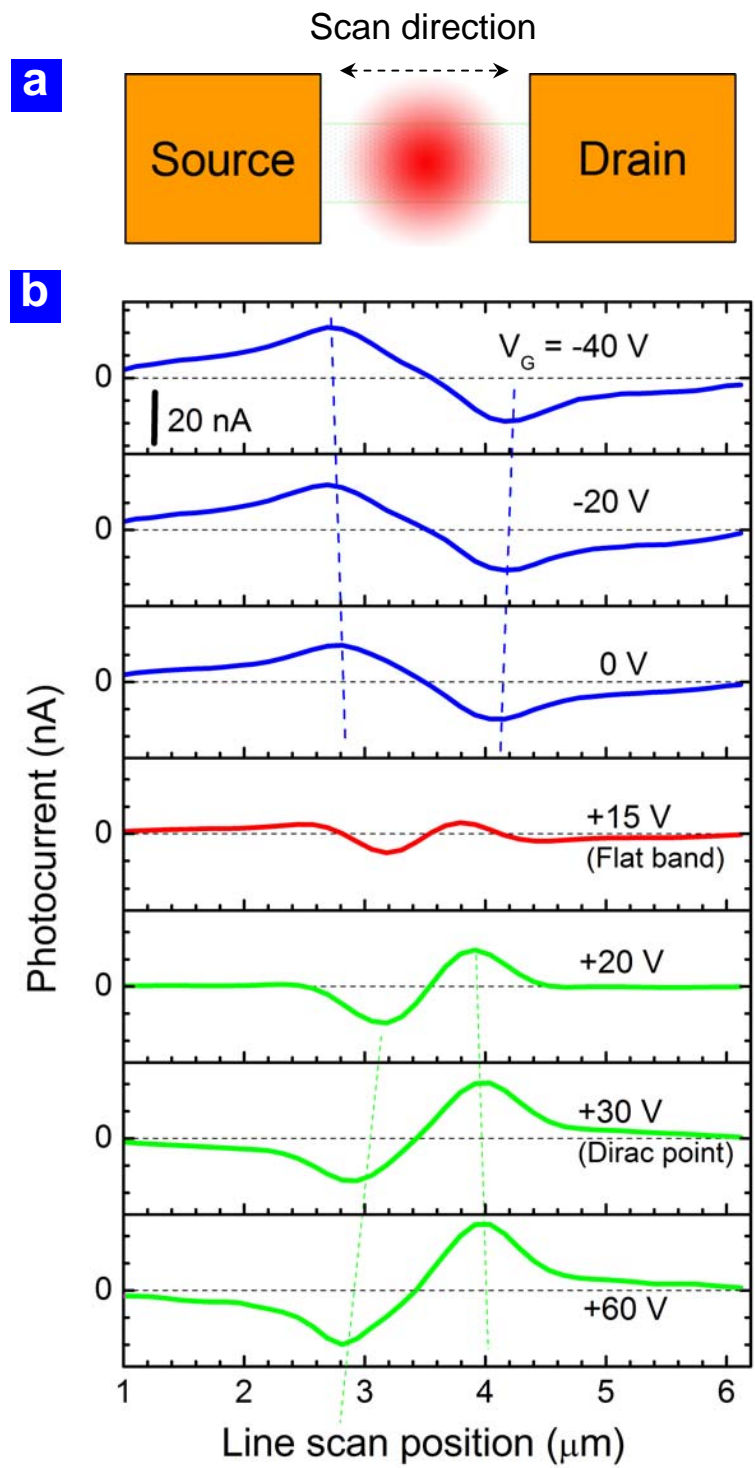

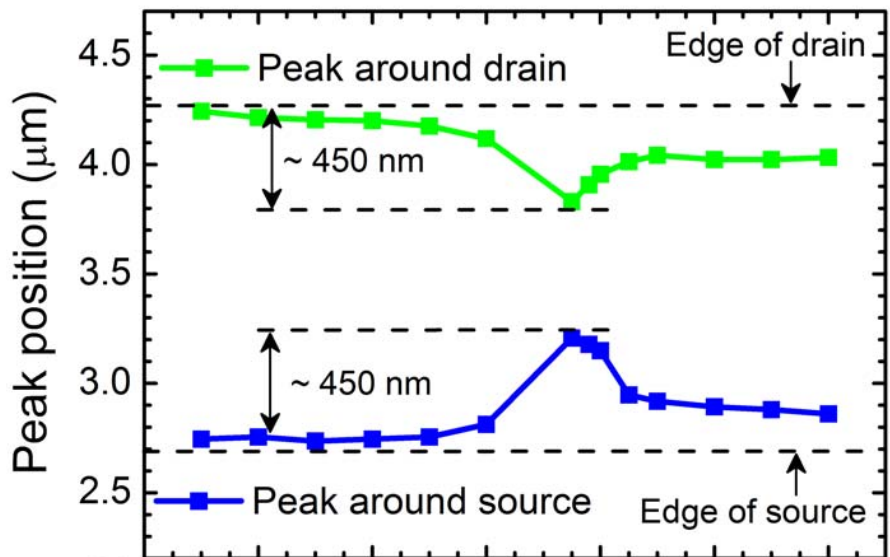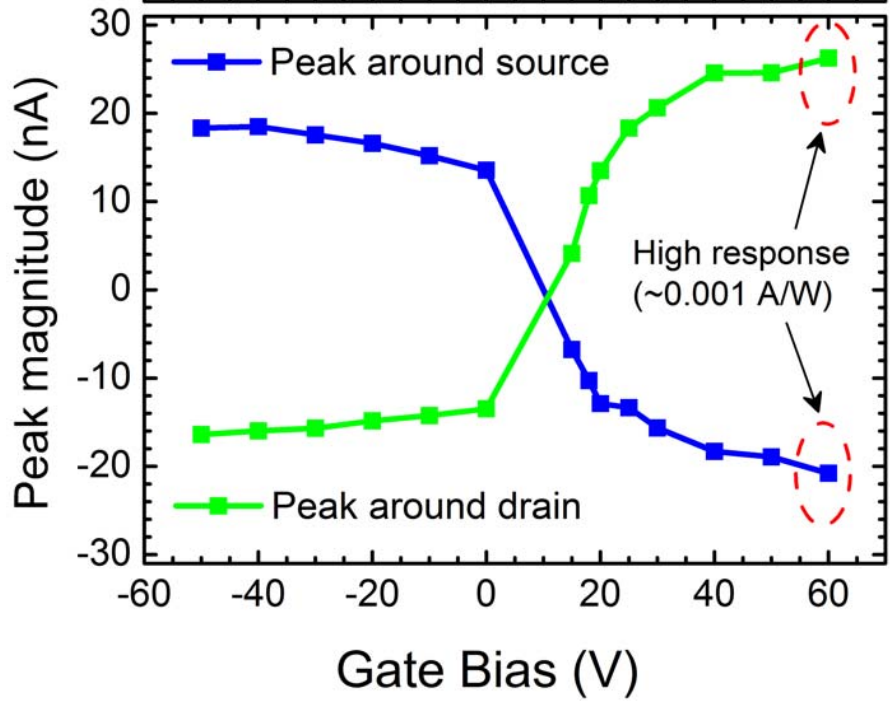

Figure 3

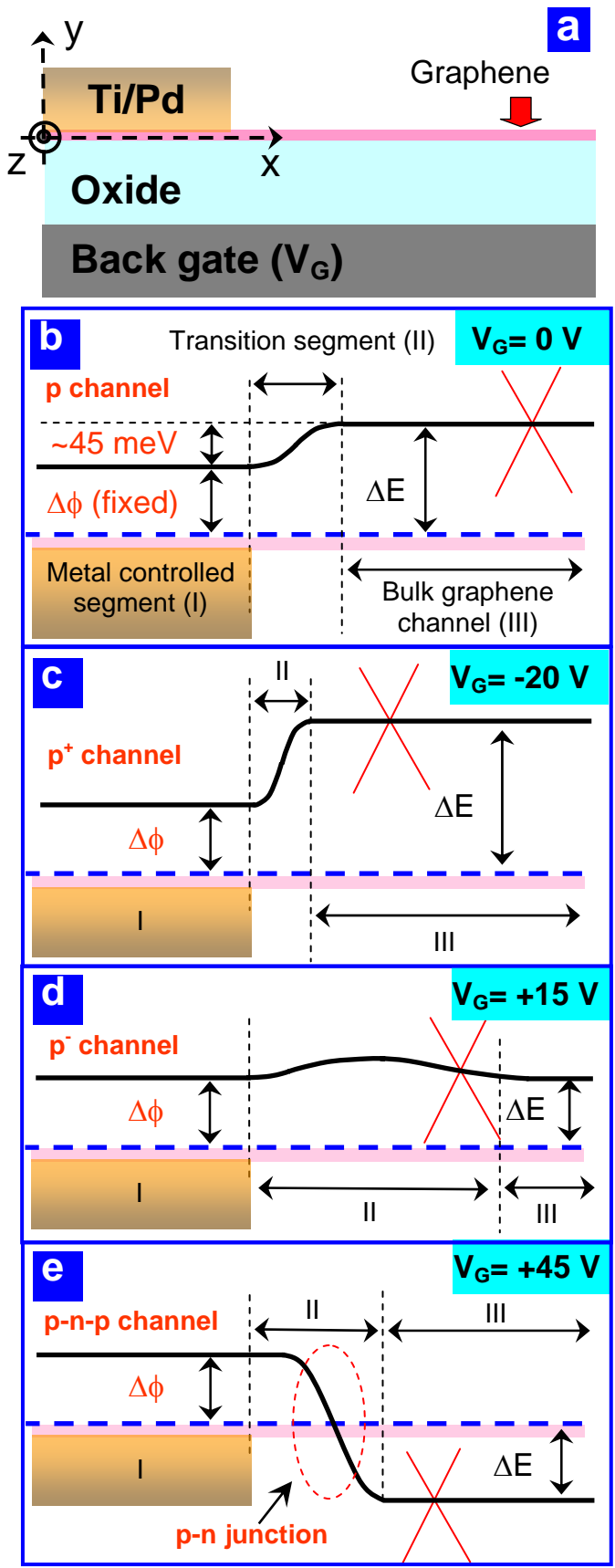

Figure 4

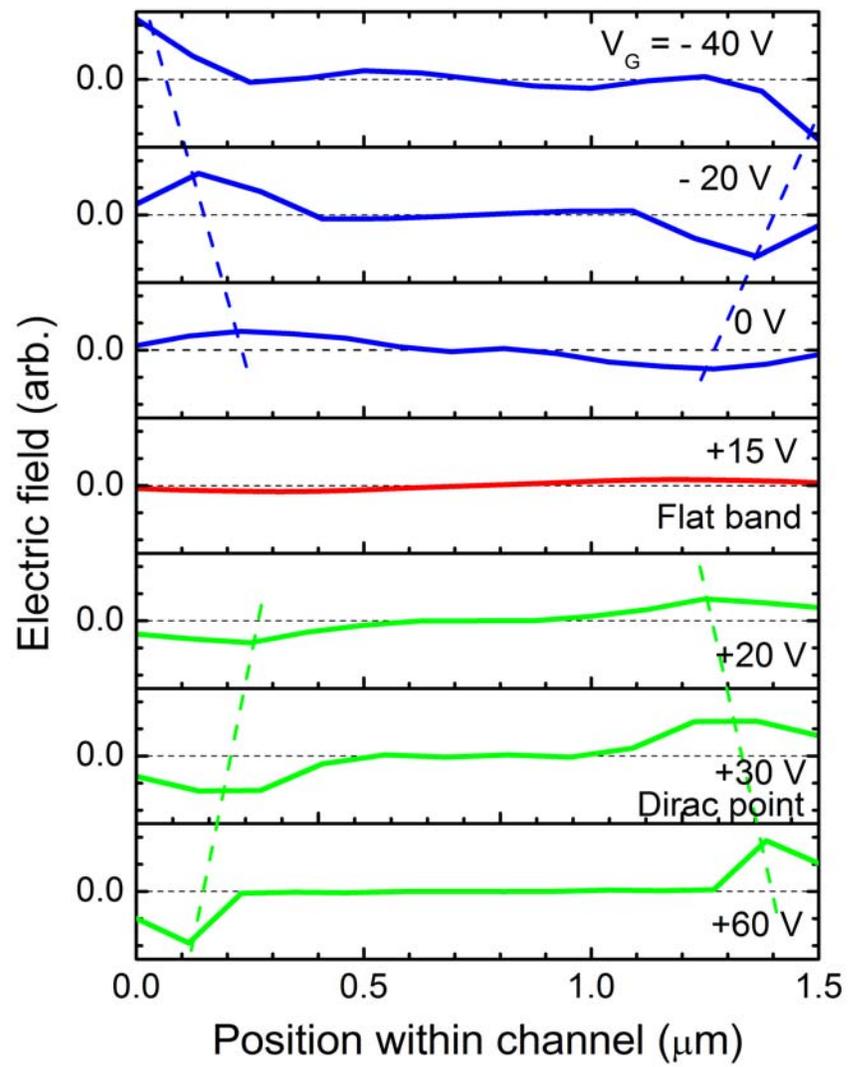

# Supporting information for "photocurrent imaging and efficient photon detection in a graphene transistor"

Fengnian Xia[1]*, Thomas Mueller[1], Roksana Golizadeh-mojarad[2], Marcus Freitag[1], Yu-ming Lin[1], James Tsang[1], Vasili Perebeinos[1], and Phaedon Avouris[1]*

[1]IBM Thomas J. Watson Research Centre, Yorktown Heights, New York 10598

[2]School of Electrical and Computer Engineering, Purdue University, West Lafayette, Indiana, 47906

*email: fxia@us.ibm.com, avouris@us.ibm.com

## I. Fabrication of graphene transistors

Fabrication of the graphene transistor began with mechanical exfoliation of Highly Oriented Pyrolytic Graphite (HOPG). Few layers of exfoliated graphenes were then deposited on top of the p+ doped substrate with 300nm thick silicon oxide and pre-patterned metallic alignment marks. Atomic force microscopy was used to determine the thickness of the graphene layers. Graphene layers with thickness of ~1 nm were further characterized by Raman spectroscopy to confirm the presence of individual graphene sheets. The transistors were produced using only one aligned e-beam lithography step, which defines the locations of the source and drain contacts. Ti/Pd/Au (1/20/20 nm) films were then deposited using e-beam evaporation, which following lift-off, formed the source and drain contacts. Some unused graphene flakes around the graphene transistor channel are also visible in the inset of Fig. 1a in the main text.

## II. Photocurrent imaging measurement

A Helium-Neon laser (632.8nm) was used as the excitation light source in the photocurrent imaging measurement. The laser beam is focused on the device using a microscope objective. The full-width-half-maximum (FWHM) of the focused laser spot is around 550 nm. A piezo-electrically driven mirror, mounted before the microscope objective, rasters the beam across the sample. The results reported in this paper were taken with a total incident optical power of ~30 µW. The incident light is modulated using a chopper at a chopping frequency of ~100 Hz. The photocurrent signal is first converted to a voltage using a low-noise current pre-amplifier and finally detected with a lock-in amplifier. In all the photocurrent measurements reported here, source and drain are always shorted ($V_S=V_D=0$ V) and the gate dependence of photocurrent ($V_G$) is studied.